\documentclass[aps,prc,reprint,groupedaddress]{revtex4-1}
\pdfoutput=1
\usepackage{graphicx}
\usepackage{epstopdf}

\begin{document}

\title{New libraries for simulating neutron scattering in dark matter detector calibrations}

\author{Alan E. Robinson}
\affiliation{Kavli Institute for Cosmological Physics and Enrico Fermi Institute, University of Chicago, Chicago, Illinois 60637, USA}
\email{fbfree@uchicago.edu}

\date{January 30, 2014}

\begin{abstract}
Dark matter detectors require calibrations of their energy scale and efficiency to detect nuclear recoils in the $1-50$ keV range.  Most calibrations use neutron scattering and require MCNP or Geant4 simulations of neutron propagation through the detector. For most nuclei heavier than $^{16}$O, these simulations' libraries ignore the contribution of resolved resonances to the neutron elastic differential cross-section.  For many isotopes and neutron energies of importance to dark matter detection, this invalid assumption can severely distort simulated nuclear recoil spectra. The correct angular distributions can be calculated from the resonance parameters using R-matrix formalism. A set of neutron scattering libraries  with high resolution angular distributions for MCNP and Geant4 of $^{19}$F, $^{40}$Ar, $^{50,52}$Cr, $^{56}$Fe, $^{136}$Xe, and $^{206,207,208}$Pb is presented.  An MCNPX library for simulating the production of low-energy neutrons in the $^9$Be$(\gamma,n)^8$Be reaction is also presented. 
Example dark matter detector calibrations are simulated with the new libraries showing how detector sensitivity could be overestimated by factors of two by relying on existing MCNP and Geant4 libraries.
\end{abstract}


\maketitle

At neutron energies up to a few MeV, neutron elastic scattering is well described by optical model scattering off a nuclear potential plus scattering off resonances of excited and compound nuclear states~\cite{Atlas_of_Resonances}.  Elastic scattering cross-sections and angular distributions can be calculated using R-matrix formalism from a list of nuclear potential shape and resonance parameters.  Many modern nuclear data evaluations using the Evaluated Nuclear Data Format (ENDF-6)~\cite{ENDF-6} provide these parameters instead of pointwise elastic scattering cross-sections~\footnote{Modern ENDF evaluations can be found though Nuclear Data Services at \protect\url{http://www-nds.iaea.org}}.  For an introduction to nuclear data evaluations in ENDF-6 format, see McFarlane~\cite{ENDF_Intro}.  The MCNP and Geant4 Monte Carlo radiation transport programs require pointwise cross-section libraries that are generated by either the NJOY~\cite{NJOY} or PREPRO~\cite{PREPRO} codes from the ENDF evaluations.

Both PREPRO and NJOY calculate neutron elastic scattering cross-sections from the resonance parameters using R-matrix formalism~\cite{Blatt}, but not the differential cross-section~\cite{*[{There is a hidden option under development in NJOY2012 for calculating the differential cross-section from resonance parameters.  }][] Bob_comm}.  Instead, these codes translate the angular distribution found in File 4 of the ENDF evaluations verbatim.  For all stable nuclei lighter than $^{16}$O, the most modern ENDF/B-VII~\cite{ENDF-VII} and JENDL-4~\cite{JENDL} evaluations contain accurate angular distributions either from R-matrix calculations or from high resolution experimental data.  However, the ENDF File 4 evaluations of almost all heavier nuclei either assume isotropy, ignore the resolved resonance contributions to the angular distributions, or are based on incomplete experimental data.

The nuclear recoil response of dark matter detectors are most often calibrated using the nuclear recoils produced by neutron elastic scattering. The simulated nuclear recoil energy distribution against which detectors are calibrated can be affected in at least three ways by incorrect elastic scattering angle distributions.
\begin{itemize}
\item Any change in the recoil energy distribution at a given neutron energy is a change in the scattering angle distribution as $E_r\propto\cos\theta$.
\item The probability for low energy neutrons to propagate into the active volume of the detector can change.
\item The energy loss and diversion of neutrons in the active volume of the detector can change, affecting multiple scattering distributions.
\end{itemize}
Calibrations that rely on simulating the absolute nuclear recoil distribution~\cite{PICASSO_calib, Horn, Juan_YBe, DAMA, *[{Ref 20. in }] [] CDMS} are vulnerable to all three effects while calibrations that determine the recoil energy and rate by tagging the outgoing neutron~\cite{Barbeau, SCENE, Juan_NaI, Manzur} are only affected by changes in the multiple scattering distributions.  Some heavy nuclei used in detector construction have resolved resonances for neutron energies only below 20 keV (producing nuclear recoils at $<1$ keV), including $^{127}$I, $^{133}$Cs, W, and most isotopes of Xe~\cite{Atlas_of_Resonances}.  These neutron recoils are below the threshold of most dark matter detectors and the use of existing neutron cross-section libraries can be used in confidence.  However, most elements heavier than oxygen have resolved resonances above 100 keV that are important in simulating the response of dark matter detectors to neutron scattering.

\begin{figure*}
\includegraphics{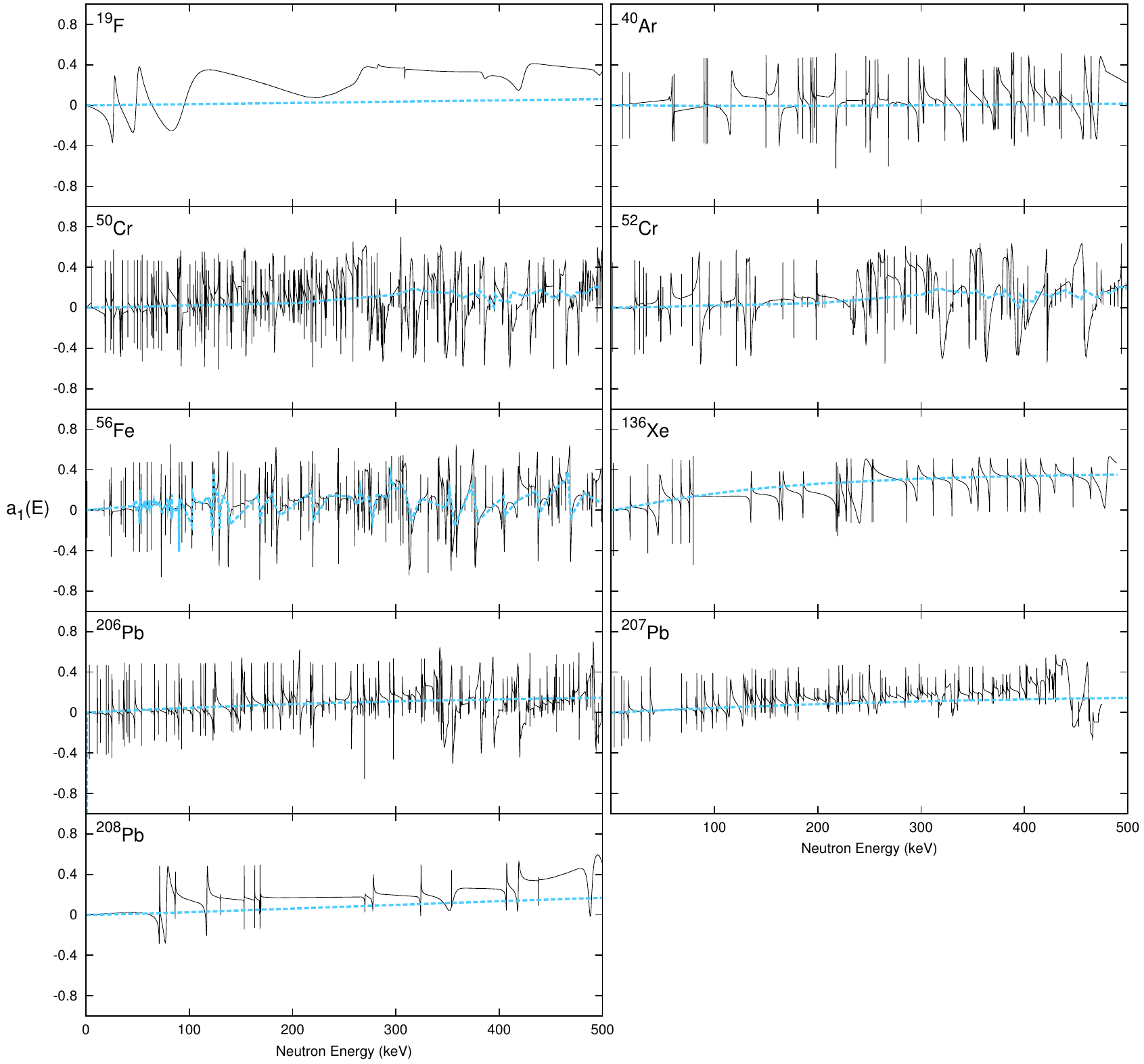}
\caption{\label{dxs_compare} Dipole anisotropy terms of neutron elastic scattering for ENDF/B-VII (dashed) and for R-matrix calculations (solid).  The R-matrix calculations are used in a set of new libraries for MCNP and Geant4.  The ENDF/B-VII iron and chromium evaluations are based on experimental data and follow the R-matrix calculated angular distribution.  The lead, xenon, and argon evaluations have angular distributions calculated using optical model calculations without resonance contributions. The $^{19}$F evaluation has no angular distribution data below 1MeV.}
\end{figure*}

Using SAMMY~\cite{SAMMY} or other R-matrix codes, the neutron scattering angular distributions can be calculated.  The SAMMY auxiliary program SAMRML can calculate the cross-section at specific angles directly from an ENDF-6 formated file.  For use in simulations, I have edited ENDF/B-VII based MCNP and Geant4~\cite{IAEA_Geant} libraries for $^{19}$F, $^{50,52}$Cr, $^{56}$Fe, $^{136}$Xe, and $^{206,207,208}$Pb with high-resolution angular distributions generated by SAMMY using R-matrix formalism.  Libraries for Si, Al, and Ge are planned.  The dipole term of the angular distributions of the ENDF/B-VII and new libraries are shown in Figure \ref{dxs_compare}.
The grids in energy and angle used by these libraries were selected to reproduce the calculated differential cross-section to better than 1\% except for $^{50}$Cr for which a 5\% tolerance was adopted.  The total memory usage of the MCNP libraries was increased by 76\% as compared to the same libraries for ENDF-VII.  To investigate the effect of the new libraries, simulation of the response of dark matter detectors to low energy neutrons with ENDF-VII.0 and these new libraries were compared.

\begin{figure}
\includegraphics{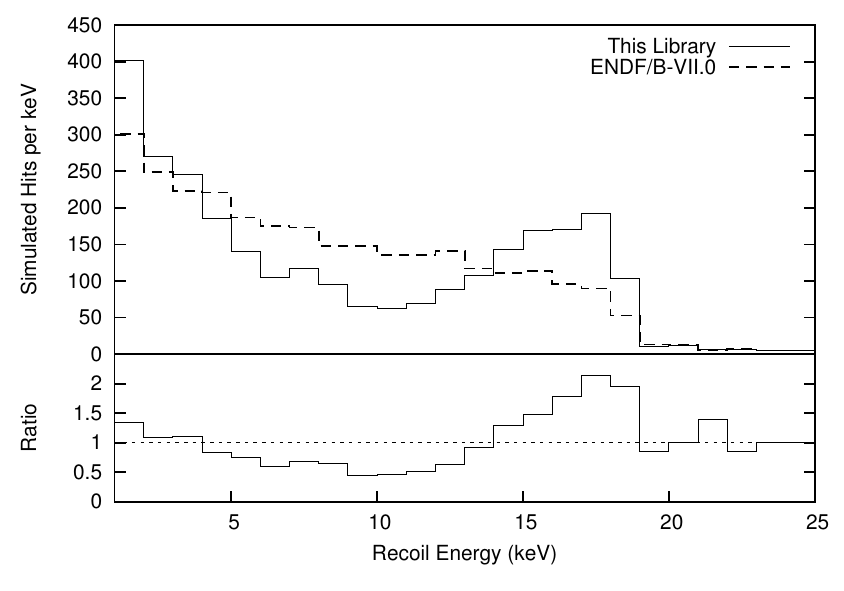}
\caption{\label{PICO_calibration} Simulated nuclear recoil distributions from 97 keV neutrons on C$_3$F$_8$ in the PICO-0.1 bubble chamber calibration experiment.  The calibration of the detector's bubble nucleation efficiency depends critically on the number of recoils at the endpoint of the simulated nuclear recoil distribution.  A factor of 2 discrepancy is found between the R-matrix calculation used in this library release and ENDF/B-VII.}
\end{figure}

These new libraries are being used by the now merged PICASSO~\cite{PICASSO} and COUPP~\cite{COUPP} (PICO) collaboration to study the response of fluorinated superheated fluid detectors.  There is an ongoing calibration of C$_3$F$_8$ in the 20 mL PICO-0.1 bubble chamber using 4.8 keV to 97 keV mono-energetic neutrons at the Universit\'{e} de Montreal's EN tandem accelerator via the $^{50}$V$(p,n)^{50}$Cr reaction.  The calibration compares the rate of bubble formation to the expected rate of nuclear recoils above the detector's threshold energy to obtain the bubble nucleation efficiency as a function of recoil energy, temperature, and pressure.
As the bubble formation rate is a convolution of the nuclear recoil energy spectrum and the bubble nucleation efficiency, the efficiency function is measured by setting the threshold energy right below the endpoint of the nuclear recoil spectrum and producing bubbles from a single known recoil energy.  The neutron energy is then changed while keeping temperature and pressure constant, and the efficiency function is deconvolved from the nuclear recoil energy spectrum.  This deconvolution is very sensitive to the measured efficiency at the recoil spectrum endpoint.
Figure \ref{PICO_calibration} shows the simulated nuclear recoil energy spectrum for a 97keV neutron beam using the ENDF/B-VII evaluation and R-matrix calculations.  At a 15keV threshold, the ENDF/B-VII evaluation over-predicts the bubble nucleation efficiency at the endpoint by a factor of 2.

These new libraries affect multiple scattering distributions by several mechanisms.  With an increase in the number of forward scatters, the neutron loses less energy at each interaction and travels further in both total track length and distance from the origin.  A simulation of 900 keV to 1 MeV neutrons propagating an infinite volume of C$_3$F$_8$ have 8\% greater track length and travel 16\% further from the origin with the new F-19 library than with the ENDF/B-VII libraries. 
The effect of the new libraries on multiple scattering distributions will depend on a detector's particular geometry and energy threshold.  The simulated probability of detecting a multiple scatter may either increase (due to more collisions) or decrease (due to particles passing through or recoils falling below threshold).

\begin{figure}
\includegraphics{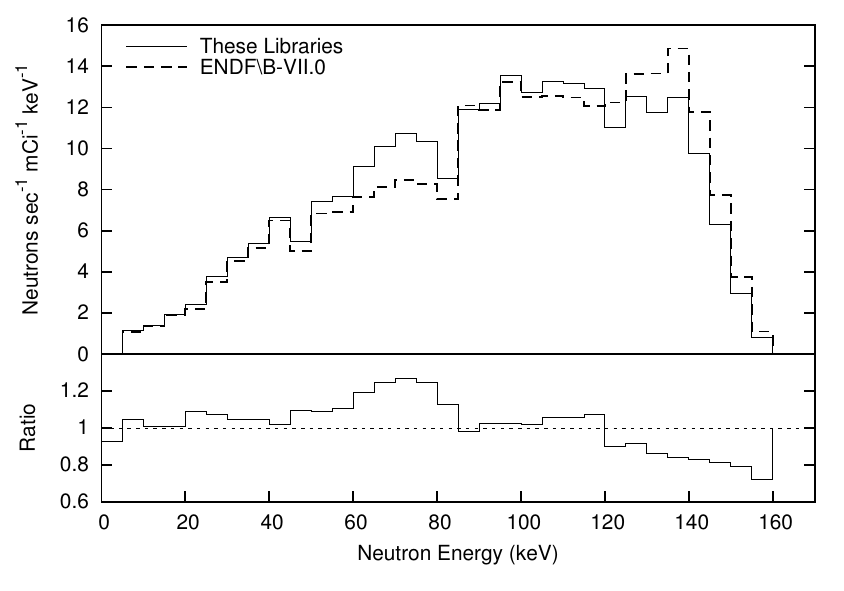}
\caption{\label{XCD_simulation} Simulated neutron energy spectra from an $^{88}$Y/Be surrounded by 20 cm of lead as used in the ongoing XCD experiment at Fermilab (see text).}
\end{figure}

XCD is a new experiment ongoing at Fermilab to calibrate a liquid xenon TPC with low energy neutrons using neutrons from an $^{88}$Y/Be neutron source, as described by Collar~\cite{Juan_YBe}.  The 152 keV neutrons from the $^9$Be($\gamma,n$)$^8$Be reaction propagate through a large amount of lead, steel, and PTFE before interacting in the liquid xenon detector.  The hit rate in the detector will be compared against the expected number of nuclear recoils from a simulation of the neutron propagation, similar to the PICO-0.1 calibration.  As with PICO-0.1, this calibration is more sensitive to high energy neutrons that are able to produce higher energy recoils in the active volume.
A simulation of the neutron energy from the lead surrounded $^{88}$Y/Be source, Figure \ref{XCD_simulation},  indicates a 17\% reduction in the number of neutrons above 130 keV exiting the lead using the new libraries as compared to using the ENDF/B-VII based libraries.

One additional new MCNPX library of the $^9$Be($\gamma,n$)$^8$Be reaction is provided in this package to allow simulations of ($\gamma,n$) neutron sources for XCD, PICO, and similar experiments.  The library implements the measured resonance parameters and branching ratios for the reaction from Arnold et al.~\cite{BeOxs} up to a maximum energy of 5.2 MeV.  A $^{88}$Y/Be source has a 5\% dipole anisotropy in both the lab-frame neutron energy and angle when converting from isotropic neutron production in the center-of-mass frame.  This anisotropy cannot be correctly coded into a spatially extended MCNP neutron source.  This library is required by MCNPX in order to obtain the correct energy-angle relationship of the neutrons.

In conclusion, a package of libraries for the simulation of low energy neutron propagation in dark matter detectors with MCNP and Geant4 is presented.  These libraries can dramatically change, by factors of 2 in some instances, the results of simulations of detector calibrations as compared to the use of presently available libraries.  The difference is especially apparent for $^{19}$F and is present at neutron energies above 20keV for all stable isotopes with $16 < A < 67$ and some heavier isotopes.

%
%
%
%

\begin{acknowledgments}
I thank my advisor J.I. Collar, C.E. Dahl for spurring this investigation, and N.E. Fields for useful discussions.  This work is supported by an NSERC Postgraduate Scholarship. Additional support was received from the Kavli Institute for Cosmological Physics at the University of Chicago through grant NSF PHY-1125897, and an endowment from the Kavli Foundation and its founder Fred Kavli.
\end{acknowledgments}

\bibliography{ref}

\end{document}